\documentclass[traditabstract]{aa}
\usepackage{txfonts}
\usepackage{graphicx}
\usepackage{float}
\usepackage{lscape}
\usepackage{natbib}
\usepackage{url}
\bibpunct{(}{)}{;}{a}{}{,}

\newcommand{\Teff}{\mbox{$T_{\rm eff}$}}

\begin{document} \title{Radiative transfer in circumstellar disks}
\subtitle{I. 1D models for GQ~Lupi} \author{S.~D. H\"ugelmeyer\inst{1}
\and S. Dreizler\inst{1} \and P.~H.~Hauschildt\inst{2} \and
A.~Seifahrt\inst{1} \and D.~Homeier\inst{1} \and T.~Barman\inst{3}}

\institute{Institut f\"ur Astrophysik, Georg-August-Universit\"at
G\"ottingen, Friedrich-Hund-Platz 1, 37077 G\"ottingen, Germany \and
Hamburger Sternwarte, Gojenbergsweg~112, 21029 Hamburg, Germany \and
Lowell Observatory, 1400 W Mars Hill Rd, Flagstaff, AZ 86001, USA}

\date{Received 17 December 2008 / Accepted 10 March 2009}


\abstract{We present a new code for the calculation of the 1D
structure and synthetic spectra of accretion disks. The code is an
extension of the general purpose stellar atmosphere code
\texttt{PHOENIX} and is therefore capable of including extensive lists
of atomic and molecular lines as well as dust in the calculations. We
assume that the average viscosity can be represented by a critical
Reynolds number in a geometrically thin disk and solve the structure
and radiative transfer equations for a number of disk rings in the
vertical direction. The combination of these rings provides the total
disk structure and spectrum.  Since the warm inner regions of
protoplanetary disks show a rich molecular spectrum, they are well
suited for a spectral analysis with our models. In this paper we test
our code by comparing our models with high-resolution VLT CRIRES
spectra of the T~Tauri star GQ~Lup.}


\maketitle

\section{Introduction}
Gas and dust disks are common objects; they can be observed around a
variety of objects such as very young stars (e.~g.\ T~Tauri and Herbig
Ae/Be stars), evolved binaries (cataclysmic variables), and even black
holes. With the discovery of the first extrasolar planets over $10$
years ago, the interest in protoplanetary disks has increased.  Disk
properties such as density, temperature, and chemical composition
effect the process of planet formation and therefore also the
characteristics of the planets. For very young stars (ages of $\sim
1-10~{\rm Myr}$), the accretion of disk matter onto the star is an
internal source of energy for the disk. The mass of such a disk is
typically a few percent of the stellar mass and mass accretion rates
are of the order of $10^{-7}-10^{-10}~M_\odot/{\rm yr}$. Furthermore,
irradiation by the central star heats the outer layers of the disk
which, in turn, radiate the impinging energy away. Therefore, the
direct detection of protoplanetary disks provides the possibility to
observe the earliest evolutionary stages of planet formation.

Tremendous efforts have been made to model the structure and more
importantly for the comparison with observations radiative transfer in
accretion disks. \citet{1986BAICz..37..129K},
\citet{1986A&A...159L...5S}, \citet{1990ApJ...351..632H} and others
went beyond the vertically averaged models of
\citet{1973A&A....24..337S} and \citet{1974MNRAS.168..603L} or models
using the diffusion approximation
\citep[e.~g.~][]{1982A&A...106...34M,1982ApJ...260L..83C} to obtain
full numerical solutions for the structure of and the radiative
transfer in accretion disks. Since then, these models have reached a
high degree of sophistication \citep[for an overview of protoplanetary
disk models see][]{2007prpl.conf..555D}.

Even though instruments like the VLTI constitute a major improvement
for the observation of spatially extended objects, most of our
information about the physics of protoplanetary disks comes from
spatially unresolved spectra. NIRSPEC observations of protoplanetary
disks have shown the capabilities of IR spectroscopy
\citep{2003ApJ...589..931N}. With the ESO IR spectrograph CRIRES
\citep{2004SPIE.5492.1218K} such observations have become available to
a larger community of astronomers. Dust continuum radiative transfer
(RT) calculations \citep[e.~g.][]{2001PhDT........11W} can reproduce
the overall structure of the dust disk out to a few hundred AU. The
more abundant gas in the disk, however, cannot be predicted by these
models. The warm inner disks (temperatures between $100~{\rm K}$ to a
few $1000~{\rm K}$) provide a special laboratory to study the gas
structure because temperatures and densities are adequate to produce
molecular spectral lines visible in the near- to
mid-infrared. High-resolution observations in combination with model
spectra enable us to obtain kinematic information about the gas since
line profiles are governed by the velocity field in the disk. Another
interesting observation is the agreement of mean inner gas disk radii
and orbital radii of short-period extrasolar planets
\citep{2007IAUS..243..135C}. To further investigate this and other
phenomena, detailed gas and dust models of the warm inner disk regions
are necessary.

Our paper is structured as follows: In Section~2 we will explain the
construction of our model structure and synthetic spectra. We will
concentrate on the basic concepts and highlight the implementation of
the solution. In Section~3 we will present synthetic spectra for the
disk of the T~Tauri star GQ~Lup and compare these with CRIRES infrared
observations to show the potential of our 1D disk model
code. Section~4 will give a summary and an outlook of our work.

\section{Models}
We have developed a circumstellar disk radiative transfer code as an
extension of the well-tested model atmosphere package \verb!PHOENIX!
\citep{jcam}. \verb!PHOENIX! can handle very large atomic and
molecular line lists and blanketing due to several hundred million
individual lines is treated in the direct opacity sampling
method. Dust is included in the models presented here by treating
condensate formation under the assumption of chemical equilibrium and
phase-equilibrium for several hundred species. I.~e.\ all dust
monomers exceeding the local saturation pressure as defined by thermal
equilibrium are allowed to condense \citep[this implies a
supersaturation ratio $S=1$;][cf.\ also
\citeauthor{2008MNRAS.391.1854H} \citeyear{2008MNRAS.391.1854H} for a
comparison of different condensation treatments]{LimDust}.

\begin{figure}[!ht]
  \centering
  \resizebox{\hsize}{!}{\includegraphics[bb=18 9 819
    540,clip]{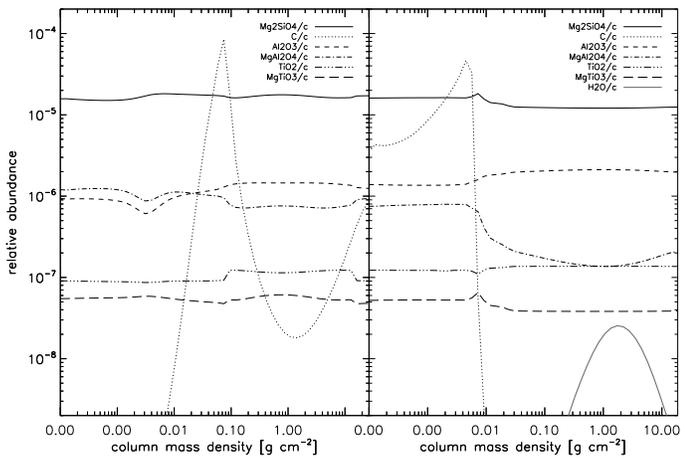}}
  \caption{Relative abundances of the most important dust
    contributors to the opacity. The left plot is for the disk ring
    with $R=0.094$~AU and the right one for $R=0.290$~AU. Forsterite
    (Mg$_2$SiO$_4$) is the most abundant dust species and strongly
    present in all layers. Graphite sets on around $m=0.01~{\rm
      g/cm^2}$ where it blocks stellar irradiation efficiently and
    therefore suppresses heating of deeper layers.}
  \label{fig:dust_abun}
\end{figure}

The grain opacity is calculated from the most important refractory
condensates using optical data for a total of 50 different
species. Absorption and scattering are calculated in the Mie
formalism, assuming a given particle size distribution for a mixture
of pure spherical grains, following the general setup of the Dusty set
of \verb!PHOENIX! atmosphere models \citep[][a plot with relative
abundances of the most important species in our disk models is shown
in Fig.~\ref{fig:dust_abun}]{LimDust}.  This equilibrium assumption is
a good approximation in the inner optically thick layers of the
well-mixed disk atmosphere. In the low-density outer layers,
non-thermal effects are more important; including these effects in the
\verb!DISK! version of \verb!PHOENIX! is planned for the future. In
the later phases of disk evolution grain growth will become important
and may cause departures from a homogeneous size distribution.
Partial pressures for the different atoms, molecules, and dust species
are calculated with the new ``Astrophysical Chemical Equilibrium
Solver'' (ACES) equation of state (Barman, in preparation) which
allows us to reach temperature regimes as low as a few tens of degrees
Kelvin.

\begin{figure}[!ht]
  \centering
  \resizebox{\hsize}{!}{\includegraphics[bb=70 45 495
    240,clip]{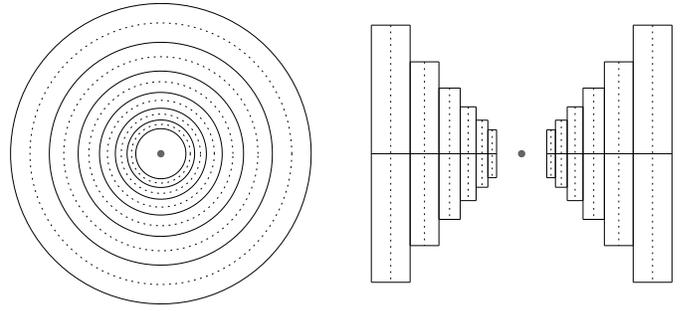}}
  \caption{\label{fig:rings} Disk ring structure as adopted for our
    calculations. The radius of the rings increases exponentially. The
    left panel shows a face-on view of a disk, the right one a
    vertical cut viewed edge-on (height is not to scale). The dotted
    lines are the radii $R$ for which the models are calculated while
    the solid lines show the borders of a disk ring. The disk
    structure is assumed to be constant over the ring width.}
\end{figure}

In our models, the disk region considered is divided into rings (see
Fig.~\ref{fig:rings}) and a plane-parallel disk atmosphere is computed
between the midplane and the top of the disk for a given number of
quadrature points (Gaussian angles) $\mu=\cos{\theta}$ for each ring
independently. Here $\theta$ denotes the angle between the
characteristic and the normal to the disk plane.

We adopt the standard accretion model for geometrically thin disks,
i.~e.\ the disk height $H$ is much smaller than the disk ring radius
$R$. This assumption decouples the treatment of vertical and radial
disk structure because the vertical structure is in quasi-static
equilibrium compared to time scales for the radial motion of
gas. Matter is assumed to rotate with Keplerian velocities and viscous
shear decelerates inner and accelerates outer parts leading to
accretion of matter and outward transportation of angular
momentum. Molecular viscosity is too small to provide the observed
mass accretion rates. Thermal convection in accretion disks was
investigated by different authors using various methods which are
summarised in \citet{2007IAUS..239..405K}. The results show that
thermal convection is unlikely the dominant source of turbulence and
even leads to inward transport of angular momentum. Furthermore, a
heating source is required to drive the
convection. \citet{1994ApJ...427..987B} argue that convective
instabilities can only occur at temperatures $T<200$~K or $2000~{\rm
K} < T < 20\,000~{\rm K}$, i.~e.~temperature regimes which our models
do not reach (see Sect.~\ref{sec:syspec}). The magnetorotational
instability (MRI) introduced by poloidal magnetic fields in weakly
ionised disk matter \citep{1991ApJ...376..214B} is the currently
favoured origin of viscosity but its effect on the thermal structure
of the disk cannot be easily described or parametrised. Even though
temperatures $T>1000$~K are necessary to thermally ionise disk
material, cosmic ray ionisation is possible at surface densities
$\Sigma < 100~{\rm g/cm}^2$ \citep{1996ApJ...457..355G} which is true
for all our models presented below. The mean viscous dissipation is
%%%%%%
%%%%%%
often modelled as an ``alpha-viscosity'' resulting in a
vertically-averaged viscosity
\begin{equation}
  \label{eq:vis}
  \bar{\nuup}=\alpha c_{\rm s} H
\end{equation}
\citep{1973A&A....24..337S} where $0 \le \alpha \le 1$ is the angular
momentum transfer efficiency, $c_{\rm s}$ the sound speed, and $H$ the
pressure scale height. Alternatively, a turbulent viscosity can be
modelled assuming a critical Reynolds number $Re$
\begin{equation}
  \label{eq:vismod}
  \bar{\nuup}=\frac{\sqrt{GM_\star R}}{Re} \quad ,
\end{equation}
\citep{1974MNRAS.168..603L}. This second model has the advantage, that
the calculation of the mean viscous dissipation is decoupled from the
thermal structure of the disk and is adopted here. Both of these
formalisms allow one to account for the effect of viscosity on the
disk structure without the need to describe its origin in detail.

\subsection{Start models}\label{sec:startmodel}
In order to start a new \verb!DISK! disk calculation either an
existing model can be used as input or a grey LTE start model is
constructed. In the latter case we follow the approach of
\citet{1990ApJ...351..632H}.

The model is initialised by setting up a mass depth scale $m_i,\
i=1,\ldots,{\rm NL}$, where ${\rm NL}$ is the number of layers, which
will be kept fixed also for later calculations with the same input
parameters. The $m$ scale is equally spaced on a logarithmic grid
between the column mass at the midplane of the disk
\begin{equation}
  \label{eq:cmass}
  M_0=\frac{\dot{M}}{6\pi
    \bar{\nuup}}\left[1-\left(\frac{R_\star}{R}\right)^{\frac{1}{2}}\right]
\end{equation}
and the outer value $m_1$, which is an input parameter. This spacing
is done for ${\rm NL}-{\rm N}_{\rm in}$ points, while for the
remaining $N_{\rm in}$ layers each point $j$, $j=1,\ldots,{\rm N}_{\rm
in}$, is positioned halfway between $M_0$ and $m_{j-1}$, with
$m_{j=0}=m_{{\rm NL}-{\rm N}_{\rm in}-1}$. This is done to provide
numerical stability in the iterative process because of the steep $m$
gradient near the midplane. A typical value for ${\rm N}_{\rm in}$ is
$6$ or $12$ for ${\rm NL}=64$.

In a next step, we calculate the depth-dependent viscosity 
\begin{equation}
  \label{eq:ddvisc}
  \nuup(m)=\bar{\nuup}(\zeta+1)\left(\frac{m}{M_0}\right)^\zeta \quad ,
\end{equation}
assuming a value of $\bar{\nuup}$ determined by an assumed constant
critical Reynolds number (Eq.~\ref{eq:vismod}), where $\zeta > 0$ is a
free parameter \citep{1986BAICz..37..129K}. The smaller the value of
$\zeta$, the lower the temperature. However, if irradiation is
noticeably strong, the effect of $\zeta$ on the structure has
negligible influence on the spectrum. \citet{1974MNRAS.168..603L}
argue that the Reynolds number has to be chosen to equal the critical
one for the onset of turbulence. We usually take $Re=10^4-10^5$ which
leads to $\alpha=0.1-0.01$.

We further assume an isothermal density structure -- the sound speed
associated with the gas pressure $c_{\rm g}$ and the flux mean opacity
$\kappa_H$ are independent of height -- which lets one derive a simple
analytic expression for the density at each depth point
$\varrho(z)$. The relation
\begin{equation}
  \label{eq:mz}
  m(z)=\int_z^0 \varrho(z) \, dz
\end{equation}
is then inverted to convert our mass depth variable $m$ into a height
above the midplane of the disk $z$.

Finally we employ the formal LTE solution derived in
\citet{1990ApJ...351..632H} and the Rosseland opacity tables of
\citet{2005ApJ...623..585F} to get a temperature structure by
iterating
\begin{equation}
  \label{eq:tvisc}
  T^4 = \frac{3}{4} T^4_{\rm eff} \left[ \tau
    \left(1-\frac{\tau}{2\tau_{\rm tot}}\right) +\frac{1}{\sqrt{3}}
    + \frac{1}{3 \epsilon \tau_{\rm tot}}
    \frac{\nuup}{\bar{\nuup}}\right]
\end{equation} 
with $\kappa_R=\kappa_R(T,\varrho)$ until the $T$-$\tau$-structure is
consistent for each layer. In Eq.~\ref{eq:tvisc} we set
$\epsilon=\kappa_R/\kappa_B$ equal to $1$ and $\tau_{\rm tot}$ is the
optical thickness at the midplane of the disk \citep[see Sec.~III b
of][]{1990ApJ...351..632H}.

\subsection{Iterative procedure}
After a restart model is read in or a start structure has been
computed, the hydrostatic equation, the radiative transfer (RT)
equation, and the energy balance equation have to be solved. This is
done iteratively until convergence in flux is obtained. Our
termination criterion is
\begin{equation}
  \label{eq:term}
  \max_{i=1,\ldots,{\rm NL}}\left|\frac{F_{\rm m}(m_i)-F_{\rm
        c}(m_i)}{F_{\rm m}(m_i)}\right| < 10^{-3} \quad ,
\end{equation}
where $F_{\rm m}$ is the expected ``mechanical'' flux released by the
disk and $F_{\rm c}$ the current flux value. This convergence
criterion in Eq.~\ref{eq:term} is not always reached and for a
sufficiently small $\max_{i=1,\ldots,{\rm NL}}\left|\Delta T_i\right|
\ll 1~{\rm K}$ the calculation is then stopped after a given number of
iterations. Typical errors in $F$ are of the order of $10^{-2}$.
\begin{figure*}[!ht]
  \centering
  \resizebox{\hsize}{!}{\includegraphics[bb=15 5 963 325,clip,
    width=17cm]{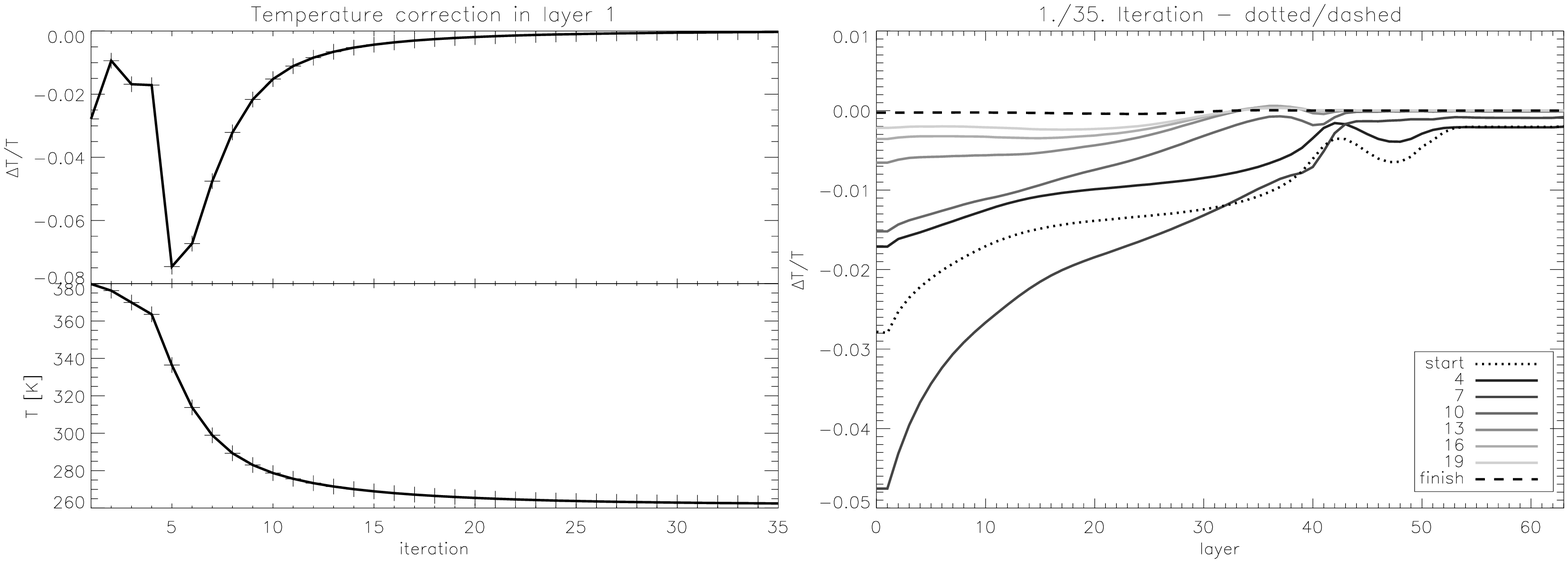}}
  \caption{Temperature correction scheme for a not irradiated disk
    atmosphere with $R=0.069~{\rm AU}$, $\dot{M}=5\cdot
    10^{-9}~M_\odot/{\rm yr}$, $Re=5\cdot 10^4$, $R_\star=1.57~R_\odot$,
    and $M_\star=0.6~M_\odot$. The top plot of the left panel shows the
    relative temperature change for each iteration in the top most
    layer. The bottom plot is the temperature in that layer. The right
    plot shows relative temperature corrections as a function of layer
    number for a chosen set of iteration steps (see legend). The dotted
    line shows the correction in the first and the dashed line in the last
    iteration. The correction is limited to $\Delta T/4$ in the first four
    iterations to avoid overcorrections. In this example convergence in
    the Uns\"old-Lucy temperature correction is reached after $\sim 35$
    iterations.}
  \label{fig:tcorr}
\end{figure*}

\subsubsection{Hydrostatic equation}
The hydrostatic equation is one of the basic equations that govern the
structure of a stable disk. Since the mass of the disk is much smaller
than that of the central object, we can neglect self-gravitation of
the disk. Assuming that the radial component of the central star's
gravitation and the centrifugal forces of the rotating disk just
cancel each other out, we obtain the vertical hydrostatic equation for
a thin disk
\begin{equation}
  \label{eq:hydro1}
  \frac{dP}{dm}=\frac{GM_\star}{R^3}z \quad ,
\end{equation}
where $P$ is the sum of gas pressure $P_{\rm g}$ and radiation
pressure $P_{\rm r}$. It is now convenient to use the relation
$dz/dm=-1/\varrho$ and replace Eq.~\ref{eq:hydro1} by
\begin{equation}
  \label{eq:hydro2}
  \frac{d^2P}{dm^2}=-\frac{c_{\rm s}^2}{P}\frac{GM_\star}{R^3} \quad .
\end{equation}
This way, we eliminate the height $z$ (which is not known a-priori)
and introduce the sound speed $c_{\rm s}^2=P/\varrho$, which is taken
from the previous iteration and only depends on the temperature
$T$. Therefore, the error in $P$ and $\varrho$ cancel out and we
obtain a more stable version of Eq.~\ref{eq:hydro1}. This second-order
differential equation is then decomposed into two coupled first-order
differential equations. With the inner boundary condition
$P^\prime(M_0)=0$ (symmetry at the midplane) and the outer boundary
condition $P(m_1)$ derived in \citet{1990ApJ...351..632H}, we solve
the two point boundary value problem with a simple shooting method.

\subsubsection{Radiative transfer}
We solve the plane-parallel radiative transfer equation 
\begin{equation}
  \label{eq:rt}
  \mu\frac{dI_\nu}{d\tau_\nu}=I_\nu-S_\nu
\end{equation}
for a given number of Gaussian angles $\mu_i$, $i=1,\ldots,{\rm NG}$,
with corresponding quadrature weights $a_i$ from the midplane ($z=0$,
$m=M_0$) to the top of the disk ($z=z_{\rm max}$, $m=m_1$). The upper
boundary condition for Eq.~\ref{eq:rt} is
\begin{equation}
  \label{eq:ubc}
  I_\nu(\nu,-\mu,z_{\rm max})=I_\nu^{\rm ext}(\nu,-\mu,z_{\rm max})
\end{equation}
and the lower boundary condition due to symmetry conditions is
\begin{equation}
  \label{eq:lbc}
  I_\nu(\nu,-\mu,0)=I_\nu(\nu,\mu,0) \quad .
\end{equation}
In Eq.~\ref{eq:ubc}, the expression $I_\nu^{\rm ext}(\nu,-\mu,z_{\rm
max})$ denotes the impinging radiation from the central star at the
top of the disk atmosphere. The radiation field $J_\nu$ is computed
using the Accelerated Lambda Iteration \citep[ALI; see,
e.~g.,][]{jcam}. Hence, we have to incorporate the boundary conditions
given in Eqs.~\ref{eq:ubc} and \ref{eq:lbc} into the formal solution
(FS) of the transfer equation (\ref{eq:rt}) and into the Approximate
Lambda Operator (ALO) $\Lambda^\star$, which is introduced to improve
the convergence rate of the Lambda Iteration by reducing the
eigenvalues of the amplification matrix. The FS can be written as
\begin{equation}
  \label{eq:li}
  J_{\rm new} = \Lambda S_{\rm old}, \quad S_{\rm
    new}=(1-\epsilon)J_{\rm new}+\epsilon B \quad ,
\end{equation}
where $S$ is the source function and $\epsilon$ is the thermal
coupling parameter. The $\Lambda$-operator is split according to
\begin{equation}
  \label{eq:opsplit}
  \Lambda = \Lambda^\star +(\Lambda-\Lambda^\star)
\end{equation}
and we can rewrite Eq.~\ref{eq:li} as
\begin{equation}
    \label{eq:ali}
    J_{\rm new} = \Lambda^\star S_{\rm
      new}+(\Lambda-\Lambda^\star)S_{\rm old} \quad .
\end{equation}
The non-local $\Lambda^\star$-operator is calculated according to
\citet{aliperf}.

\subsubsection{Energy balance}
The standard accretion disk model demands that all the energy that is
produced by viscous dissipation, i.~e.\ mechanical energy, is released
in form of radiative and convective energy, viz.
\begin{equation}
  \label{eq:eneq}
  E_{\rm m}(m) = E_{\rm r}(m) + E_{\rm c}(m) \quad .
\end{equation}
For a viscous disk, the left hand term can be written as
\begin{equation}
  \label{eq:enmch}
  E_{\rm m}(m) = \frac{9}{4}\frac{GM_\star}{R^3}\nuup(m)\varrho(m)
\end{equation}
\citep{1986BAICz..37..129K} and the radiative energy is expressed by
\begin{equation}
  \label{eq:enrad}
  E_{\rm r}(m)=4\pi\int_0^\infty \left(\eta_\nu(m)-\chi_\nu(m)
    J_\nu(m) \right) d\nu
\end{equation}
We shall neglect the convective energy term $E_{\rm c}$ from here on
because radiation is dominating the temperature structure and
convection seems to have little effect \citep{1998ApJ...500..411D}.
We then derive an Uns\"old-Lucy class temperature correction scheme
\citep{1964SAOSR.167...93L} by integrating the second
frequency-integrated moment of the radiative transfer equation
\begin{equation}
  \label{eq:moments2}
  \frac{dK}{dm} = \kappa_H H \quad ,
\end{equation}
where 
\begin{equation}
  \label{eq:kaph}
  \kappa_H=\int_0^\infty \left(\chi_\nu / \varrho\right) H_\nu \,
  d\nu / H \quad .  
\end{equation}
The introduction of the Eddington factors
\begin{equation}
  \label{eq:efs}
  f_K(m)=\frac{K(m)}{J(m)} \quad {\rm and} \quad f_H=\frac{H(0)}{J(0)}
\end{equation}
and insertion of the first frequency-integrated moment of the
radiative transfer equation gives the temperature correction law
\begin{eqnarray}
  \label{eq:ultc}
  \Delta T &=& \frac{\pi}{4 \sigma T^3}
  \biggl[ \frac{\kappa_J}{\kappa_B} \left( \frac{\Delta
          H(0) f_K(0)}{f_K f_H} + \frac{1}{f_K} \int_0^m \kappa_H
        \Delta H(m') dm'\right) - \nonumber \\
      &&\frac{1}{\kappa_B} \frac{d \Delta H(m)}{dm} \biggl] \quad ,
\end{eqnarray}
with
\begin{equation}
  \label{eq:kaps}
  \kappa_J=\int_0^\infty \left(\kappa_\nu / \varrho\right) J_\nu \,
d\nu / J \quad\mathrm{and}\quad \kappa_B=\int_0^\infty
\left(\kappa_\nu / \varrho\right) B_\nu \, d\nu / B ,
\end{equation}
the absorption mean and Planck mean opacity. An example iteration
history is shown in Fig.~\ref{fig:tcorr}.

\subsubsection{Irradiation}
\begin{figure}[!hb]
  \centering
  \resizebox{\hsize}{!}{\includegraphics[bb=249 225 1043
    515,clip]{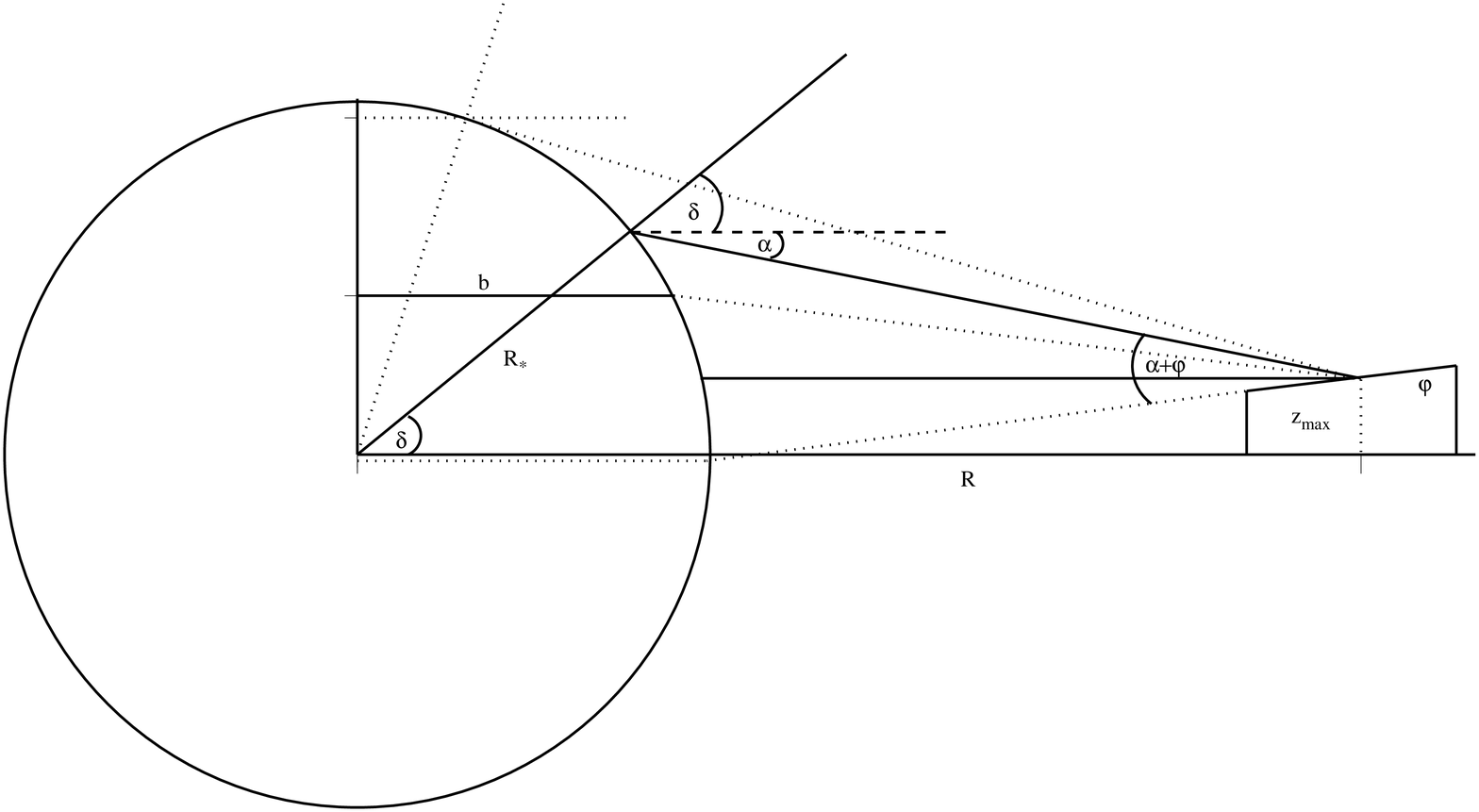}}
  \caption{Irradiation geometry as adopted for our calculations. We
    consider a star with radius $R_\star$ at a distance $R$ from a
    disk ring with a height $z_{\rm max}$ and a slope of the disk
    surface $\cos(\varphi)$. For each Gaussian angle
    $\mu_i=\alpha+\varphi$ and its corresponding integration weight
    $a_i$ the projected surface fraction on the star is determined and
    the irradiation intensity is calculated. The angle under which
    radiation leaves the surface of the star is
    $\mu_j=\alpha+\delta$.}
  \label{fig:irr}
\end{figure}

Irradiation by the central star plays an important role in the
determination of the temperature and height profile of a
protoplanetary accretion disk. Therefore, we have taken special care
to treat this effect in detail. The impinging intensity onto the
surface of the disk (see Eq.~\ref{eq:ubc}) is determined by first
calculating the slope of the disk surface $\cos(\varphi)=\Delta
z/\Delta R$ at ring radius $R$ by computing the height of the disk
according to our start model (see Section~\ref{sec:startmodel}) at
$R^\prime=R\pm\Delta R$ with $\Delta R=R \cdot 10^{-2}$. For each
Gaussian angle $\mu_i$ and corresponding integration weight $a_i$ the
projected surface fraction on the star $A_i$ is determined. This area
is then subdivided and a mean intensity reaching the disk surface is
evaluated for a star with a blackbody spectrum according to
\begin{equation}
  \label{eq:irr}
  \bar{I_\nu}(\mu_i,\nu) = \frac{B_\nu}{2\pi}
  \left( \sum_{j=1}^{\mathrm{NS}} A_j \left(\frac{R_\star}{r_j}\right)^2\frac{
      I_\nu(\mu_j,\nu)}{I_\nu(1,\nu)} \right)  \ \ \mathrm{with} \ \
  A_j=\frac{A_{\mathrm{proj}}}{\pi R_\star^2} \ ,
\end{equation}
where ${\rm NS}$ is the number of surface segments for Gauss angle
$\mu_i$, $r_j$ is the distance from the area element $A_j$ to the disk
surface, and $\mu_j$ the angle under which the radiation leaves the
star. The factor $1/2\pi$ is needed since the the irradiation onto the
disk is unidirectional and not isotropic. Limb darkening is considered
by the factor $ I_\nu(\mu_j,\nu)/I_\nu(1,\nu)$. Alternatively, a
\verb!PHOENIX!  spectrum can be used as irradiation source. The
irradiation geometry is shown in Fig.~\ref{fig:irr}.

\subsubsection{Spectrum}
Having determined a set of self-consistent disk ring structures, we
%%%%%%
%%%%%%
calculate a last iteration with a larger number of Gaussian angles
(usually ${\rm NG}=16$) and a fine wavelength spacing (up to $\Delta
\lambda=0.2~{\rm \AA}$ in the wavelength range of interest) to obtain
a high-resolution model spectrum. Since the Keplerian rotation
velocity and the inclination angle $i$ of the disk are not considered
in the single ring spectra yet, we compute a combined spectrum
according to
\begin{eqnarray}
  \label{eq:spec}
  \bar{I}_\nu(\nu,i)&=&\cos(i) \int_{R^{\rm in}}^{R^{\rm out}}
  \hspace{-7pt} \int_0^{2\pi} I(\nu,\phi,r,i) \, r \, d\phi \, dr \\
  \label{eq:specnum}
  &=&\pi \cos(i) \sum_{j=1}^{\mathrm{NR}} \biggl\lbrace
  \left[\left(R_j^{\rm out}\right)^2-\left(R_j^{\rm
        in}\right)^2\right] \cdot  \nonumber \\
  && \hspace{1.8cm} \sum_{k=1}^{\mathrm{NA}}
  I\left(\nu^\prime(i,\phi_k),\phi_k,R_j,i\right) \biggr\rbrace\quad .
\end{eqnarray}
In Eq.~\ref{eq:specnum} we assumed that the disk is axis-symmetric and
that the intensity is constant for all radii between inner and outer
%%%%%%
\begin{figure*}[!ht]
  \centering
  \resizebox{\hsize}{!}{\includegraphics{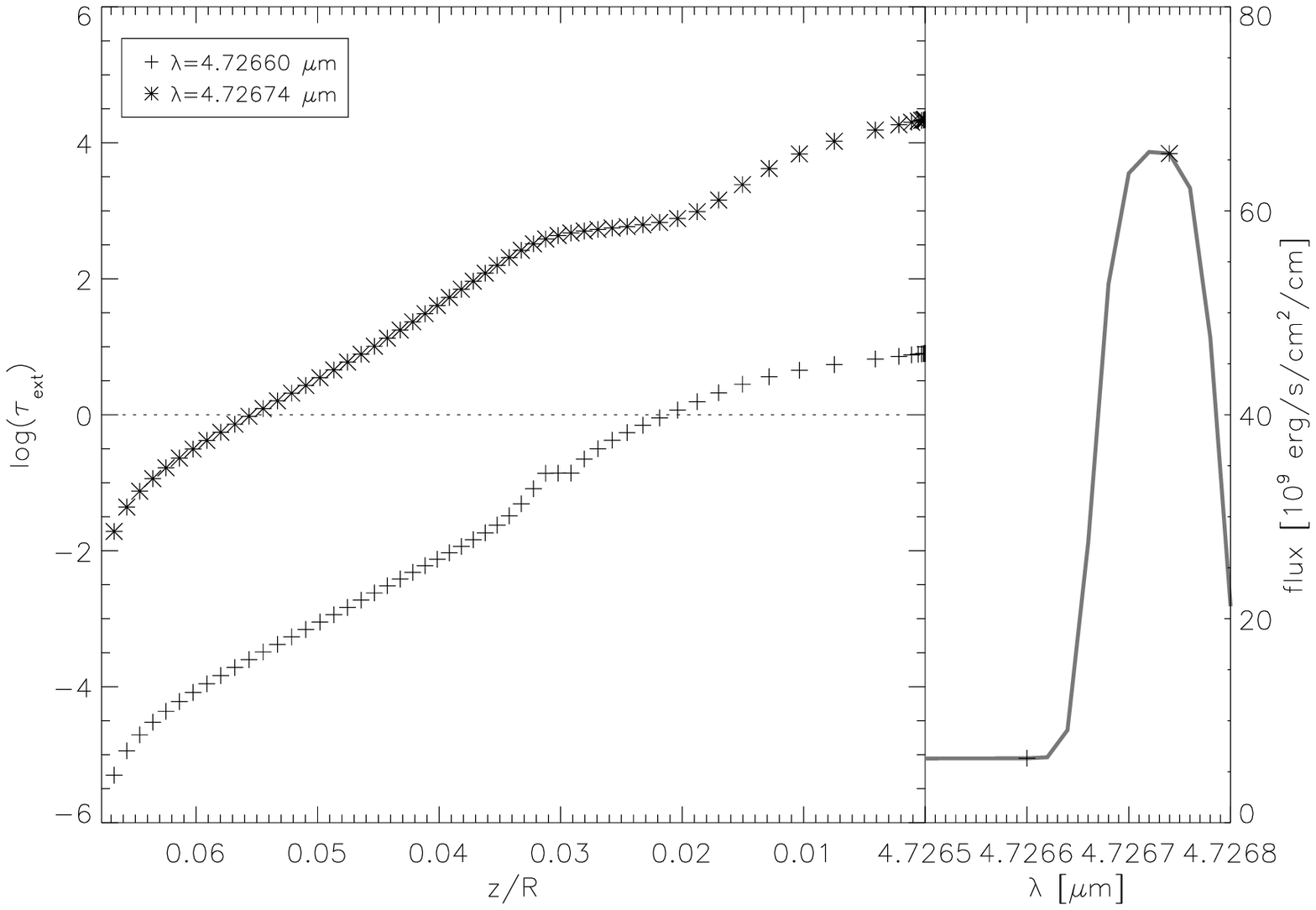}\includegraphics{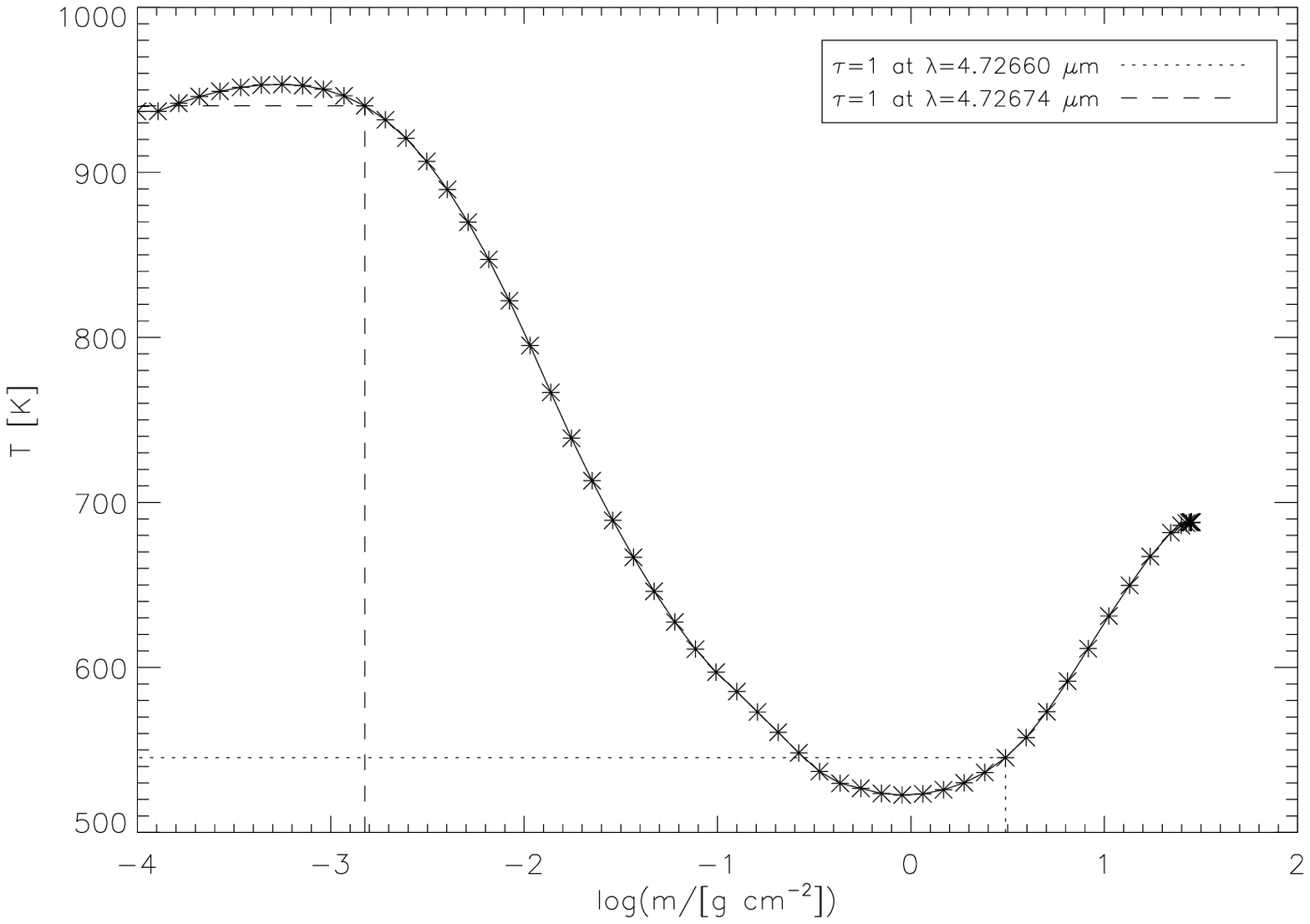}}
  \caption{Optical depth structure for a disk ring model with
    $R=0.065~{\rm AU}$, $M_\star=0.8~M_\odot$, $R_\star=2.55~R_\odot$,
    $\dot{M}=3\cdot10^{-9}~M_\odot~{\rm yr}^{-1}$, and $Re=5\cdot 10^4$. The
  left plot shows the run of $\tau_{\rm ext}$ with height ($z/R=0$ is
  the midplane of the disk). The asterisks denote the optical depth at
  the center of the line, the crosses are for a continuum point (see right
  panel). The dotted line at $\tau=1$ marks the region where the line
  and the continuum become optically thick, i.~e.\ where the
  radiation that we see comes from. The plot on the right is a
  temperature-$\log{m}$-diagram showing at which temperatures
  and column masses the optical depth for the line (dashed line) and
  the continuum (dotted line) become unity.}
  \label{fig:tau}
\end{figure*}
%%%%%%
radii for each ring. In addition to the integration over all disk
rings, the influence of the disk's rotation on the line profile is
taken into account by applying the Doppler shift
\begin{equation}
  \label{eq:doppler}
  \nu^\prime(i,\phi)=\nu
  \sqrt{\frac{1-u(i,\phi)/c}{1+u(i,\phi)/c}}
\end{equation}
to the line. Here the velocity $u(i,\phi)=u_{\rm
Kepler}\sin(i)\cos(\phi)$ is determined for a given inclination $i$
and for a set of disk ring segments with azimuthal angle $\phi$. We
use $\mathrm{NA}=100$ disk ring segments, i.~e.\ $100$ steps in $\phi$, to
determine the rotationally broadened line profile. This method is a
simple way to determine line profiles for rotating accretion disks if
the lines originate in the very upper layers of the disk. However,
\citet{1986MNRAS.218..761H} noted that an anisotropy in the local
emission pattern changes the global line shape: line photons are
trapped in optically thick emission layers and can more easily escape
in directions where there are larger Doppler gradients. The true
consequence of this on the line shape will be discussed in a future
paper (H\"ugelmeyer et al. in preparation) where we will present full
3D radiative transfer calculations in rotating accretion disks.
%%%%%%
\begin{figure}[!hb]
  \centering
  \resizebox{\hsize}{!}{\includegraphics{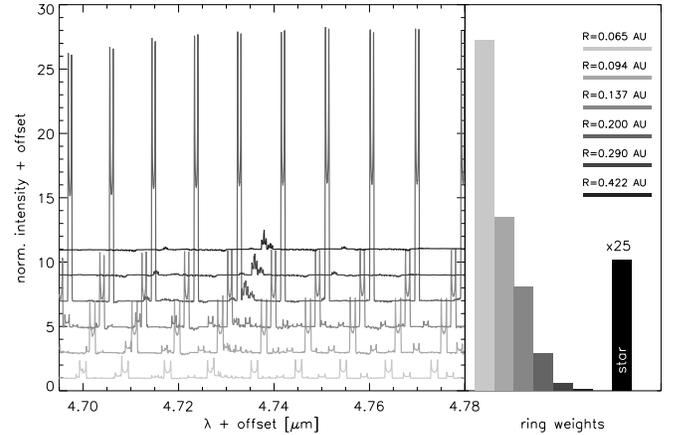}}
  \caption{The left panel shows normalised disk ring
      spectra. Intensities and wavelength are offset for clarity. The
      right panel depicts bars corresponding in height to the contribution
      of each disk ring spectrum to the total spectrum. The bar for the
      stellar contribution has to be multiplied by a factor of $25$. The
      spectra and the weights are grey-scale coded corresponding to their
      ring radius $R$.}
    \label{fig:linecont}
\end{figure}
%%%%%%
%%%%%%
\begin{figure}[!h]
  \centering
  \resizebox{\hsize}{!}{\includegraphics[bb=13 8 490
    342,clip]{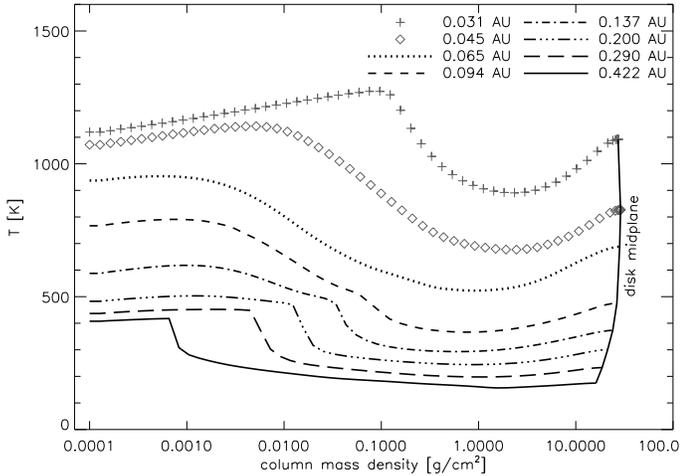}}
  \caption{Temperature structures for the eight disk rings calculated
    for the spectral analyses. The temperature profiles marked with
    symbols, e.~g. for disk rings with $R=0.031$~AU and $R=0.045$~AU,
    are not included in the best fit model.}
  \label{fig:tall}
\end{figure}
%%%%%%

\section{\label{sec:syspec}Synthetic spectra for GQ~Lup}
We retrieved spectra of T~Tauri stars taken with the high-resolution
infrared spectrograph CRIRES at the VLT from the ESO Science Archive
Facility \citep[see][for a description of the
observations]{2008ApJ...684.1323P}. The observations were reduced
using a combination of the CRIRES pipeline and our own IDL
routines. The telluric absorption lines in the spectrum were removed
by using a telluric model spectrum (Seifahrt, in preparation). From
the spectra taken between April 22 and April 26 2007, we selected the
observation of the classical T~Tauri star GQ~Lup to demonstrate the
applicability of our models to observations.

\subsection{\label{ssec:gqlup}GQ~Lup}
GQ~Lup is a classical T~Tauri star (CTTS) which is mostly known for
its recently discovered sub-stellar companion GQ~Lup~B
\citep{2005A&A...435L..13N}. The activity due to ongoing accretion of
the CTTS makes a well constrained determination of its physical
parameters difficult. This becomes obvious regarding the differences
in visual brightness of more than $2~{\rm mag}$ \citep[$V_{\rm
max}=11.33~{\rm mag}$ and $V_{\rm min}=13.36~{\rm
mag}$;][]{2006A&A...453..609J}. \citet{2007A&A...468.1039B} and
\citet{2008A&A...489..349S} (BR07 and SD08 from here on) both obtained
photometric and spectroscopic data to determine rotational periods and
$v \sin i$ for GQ~Lup~A and also to derive other stellar parameters
which we need as input for our models. Both authors assume an
effective temperature for a K7~V star of $\Teff=4060~{\rm K}$
\citep{1995ApJS..101..117K}.

SD08 obtained a radius of $R_\star=1.8\pm 0.3~R_\odot$ for the K7~V
star assuming a mean distance of $150\pm20~{\rm pc}$.  From
evolutionary tracks they derive a mass of $M_\star=0.8 \pm
0.2~M_\odot$. Together with their rotational period of $10.7 \pm
1.6~{\rm d}$ and a spectroscopically determined $v \sin i = 6.5 \pm
2.0~{\rm km~s}^{-1}$ they find an inclination angle of $i=51^\circ \pm
13^\circ$.

BR07 measure a shorter photometric period of $8.45 \pm 0.2~{\rm d}$
and a larger radius of $R_\star=2.55\pm 0.41~R_\odot$ from $T_{\rm
eff}$ and a luminosity $L$ from evolutionary tracks. With these values
and $v \sin i = 6.8 \pm 0.4~{\rm km~s}^{-1}$ they predict an
inclination angle of $i=27^\circ \pm 5^\circ$.

\subsection{Models}
We calculated small model grids for the input parameters from the
publications of BR07 and SD08 (see Sec.~\ref{ssec:gqlup}), i.~e.\
using $\Teff=4060~{\rm K}$, $M_\star=0.8~M_\odot$, and radii of
$R_\star=2.55~R_\odot$ and $R_\star=1.80~R_\odot$. The different radii
lead to luminosities of $L=1.58~L_\odot$ and $L=0.79~L_\odot$,
respectively. We varied the mass accretion rate ($\dot{M}=8\cdot
10^{-10} - 2\cdot 10^{-8}~M_\odot~{\rm yr}^{-1}$) and the Reynolds
number ($Re=5\cdot10^4 - 1\cdot10^5$). For each parameter set we
computed eight disk rings with $R=0.031,\ 0.045,\ 0.065,\ 0.094,\
0.137,\ 0.200,\ 0.290,\ {\rm and}\ 0.422~{\rm AU}$. The ring radius
$R$ increases exponentially to achieve a small temperature gradient
from ring to ring. Figure~\ref{fig:tall} shows typical temperature
structures for a set of disk rings. The decreasing temperature in the
outer layers stems from the fact that our temperature is set by the
gas which has a relatively low opacity at the wavelengths where
stellar irradiation is the strongest. This effect becomes small at
radii larger than $R=0.065$~AU.

Since we consider a disk that has recently formed from a protostellar
cloud and yet not seen significant grain growth, we assumed a standard
ISM type power law grain size distribution with base size
$a_0=6.25\cdot 10^{-3}~{\rm \mu m}$ and an exponent of $-3.5$. The
abundances follow \citet{gn93}.

In Fig.~\ref{fig:linecont} we show the normalised spectra of each disk
ring and how much each spectrum contributes to the total disk spectrum
in the wavelength region considered. From these ring weights one can
see that the outer disk radius $R_{\rm out}$ contributes only very
little to the ring integrated spectrum and that the extension to even
larger disk radii would have almost no measurable
influence. Furthermore, it becomes obvious that the inner rings, even
though their surface area is much smaller than those of rings at
larger radii, have a much higher weight. This is simply because the
irradiated and viscously heated atmosphere is much warmer closer to
the star and therefore has a higher flux level than that of rings
further out. Furthermore, the continuum becomes optically thin for
disk parts with $R \ga 0.200~{\rm AU}$ which further decreases the
flux level. The influence of the central star GQ~Lup~A on the total
spectrum is accounted for by a blackbody spectrum of $4060~{\rm
K}$. The ratio of stellar and disk flux is $F_{\rm star}/F_{\rm disk}
\sim 5$ depending on the input parameters for the
disk. Figure~\ref{fig:linecont} also shows that CO emission lines
disappear at $R\ge 0.290~{\rm AU}$ which corresponds to temperatures
in the outer line emitting layers of $T<500~{\rm K}$.

\subsection{Spectral fit}
\begin{figure*}[!ht]
  \centering
  \resizebox{\hsize}{!}{\includegraphics[bb=47 36 777
    615,clip]{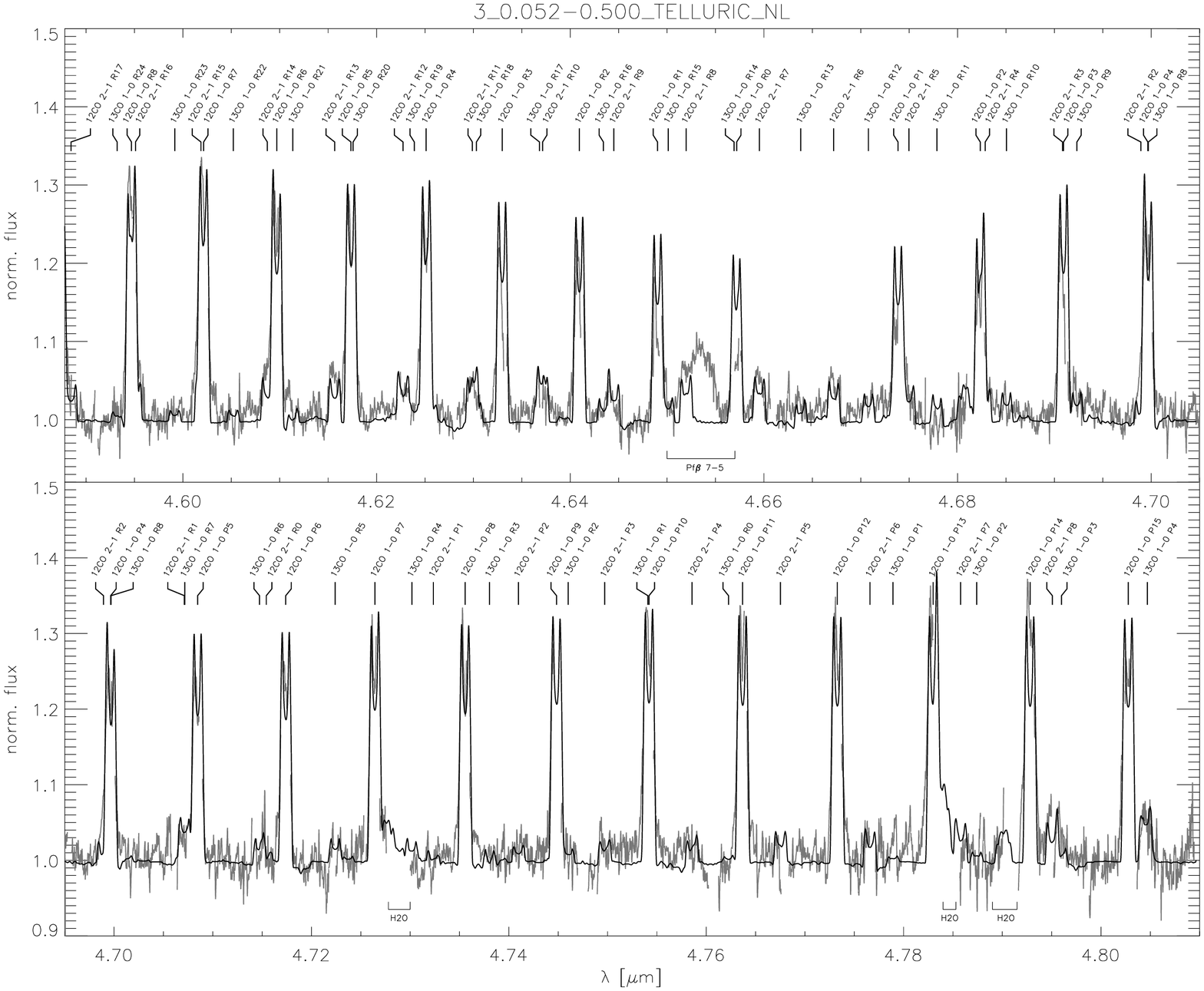}}
  \caption{CRIRES spectrum of GQ~Lup (grey line) with our best fit
    disk model (black line). The model is calculated for a disk
    between $R_{\rm in}=0.052~{\rm AU}$ and $R_{\rm out}=0.500~{\rm
      AU}$ with a mass accretion rate of $\dot{M}=3\cdot
    10^{-9}~M_\odot~{\rm yr}^{-1}$ and a Reynolds number of
    $Re=5\cdot10^4$. The stellar parameters are those of
    \citet{2007A&A...468.1039B}: $R_\star=2.55~R_\odot$ and $T_{\rm
      eff}=4060~{\rm K}$. For the stellar mass we adopted the value of
    $M_\star=0.8~M_\odot$ given by \citet{2008A&A...489..349S}. The
    hydrogen Pf$\beta$ (7$-$5) line at $4.654~\mu{\rm m}$ cannot be
    attributed to the disk because its Doppler width corresponds to a
    Kepler velocity of $R=0.025~{\rm AU}=2~R_\star$ which lies well
    within our determined inner disk radius.}
  \label{fig:gqlup}
\end{figure*}

The comparison of our disk model spectra to the CRIRES spectrum of
GQ~Lup in the wavelength range $\lambda=4.59-4.81~{\rm \mu m}$
indicates that the inclination of $i=51^{\circ} \pm 13^{\circ}$
derived by SD08 is too high: the emission lines are much broader in
the model than in the observation. Even if we set $R_{\rm in}$ to
large values $> 0.200~{\rm AU}$, i.~e.\ excluding the inner rings
which have broader lines due to larger Kepler velocities, the line
widths cannot be acceptably fitted. For the input parameters of BR07,
%%%%%%
\begin{figure}[!ht]
  \centering
  \resizebox{\hsize}{!}{\includegraphics[bb=23 8 496
    343,clip]{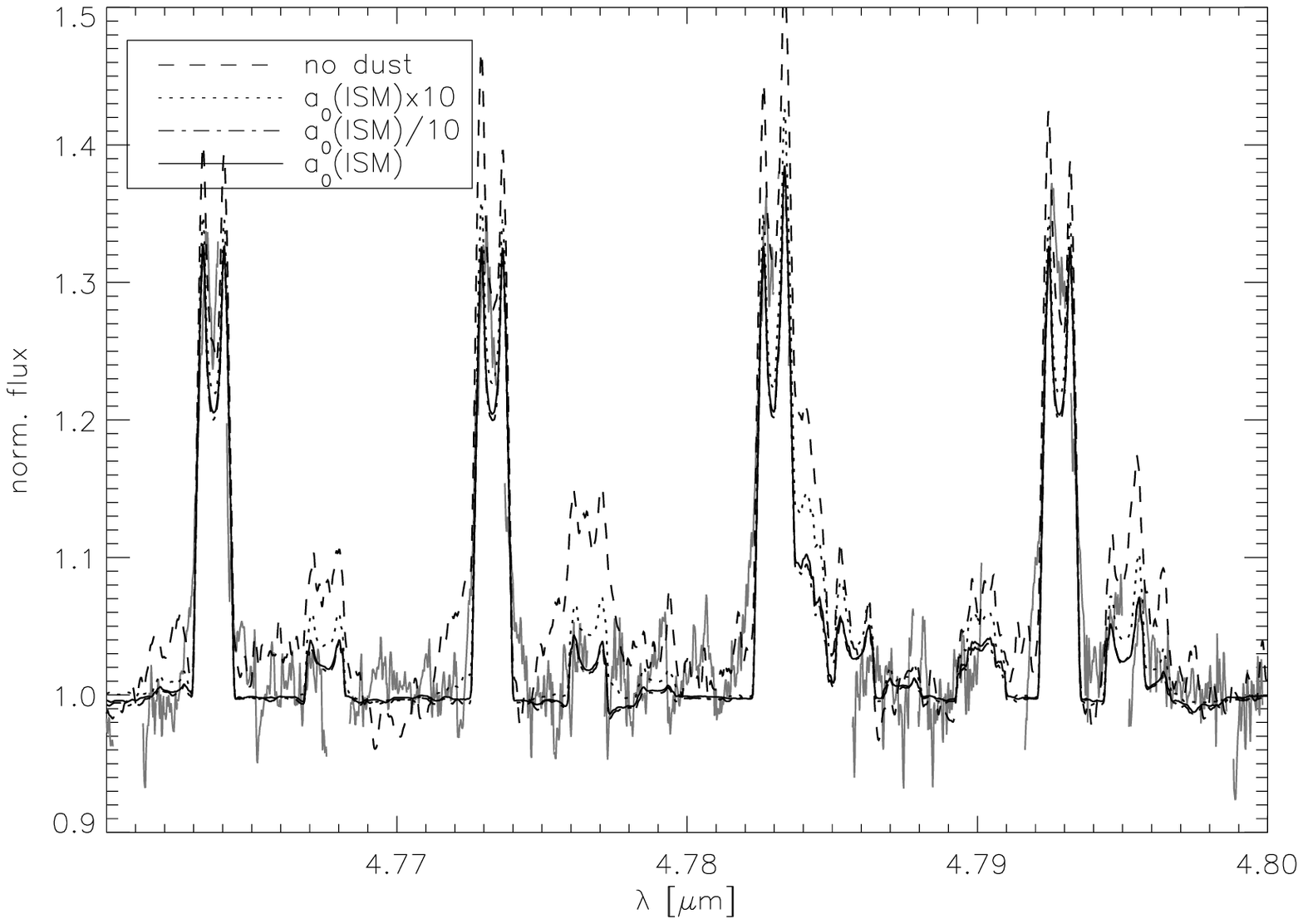}}
  \caption{Spectra calculated from the same structure model
    but for different dust grain sizes. Neglecting dust as opacity
    source results in overestimation of the lines. ISM grain size
    distribution and a dust grain size setup with ten times smaller
    dust radii yield a very similar spectrum. The dust opacity is
    slightly reduced if we increase the grain size by a factor of
    ten.}
  \label{fig:dust_spec}
\end{figure}
%%%%%%
we find a reasonably good fit to the CRIRES data (see
Fig.~\ref{fig:gqlup}). The line widths are best fitted with an
inclination angle of $i=22^\circ$, which is the lowest possible value
given by the errors in BR07. The line strengths are best reproduced if
we set the inner radius to $R_{\rm in}=0.052~{\rm AU}$ and the outer
radius to $R_{\rm out}=0.500~{\rm AU}$ for a mass accretion rate of
$\dot{M}=3\cdot 10^{-9}~M_\odot~{\rm yr}^{-1}$ and $Re=5\cdot~10^4$
corresponding to $\alpha \approx 0.05$. The line wings of the
fundamental transition lines ($v=1-0$) close to the continuum are
somewhat broader in the observation than in the model. This argues for
contributions of disk regions with high Kepler velocities, i.~e.\
smaller radii $R$. However, if we include disk rings with smaller
radii, the temperature sensitive $v=2-1$ CO emission lines are too
strong in the model. Furthermore, the low $J$ R- and P-branch CO
$v=1-0$ lines around $4.66~\mu{\rm m}$ are overestimated by the model,
or put in other words, the increment of the line strengths is too
strong in the model. For our best model fit, we use a model which fits
the strength of the higher order CO $v=1-0$ lines because these are
more frequent in the observation accepting a less well reproduction of
the low order lines.
%%%%%%
%%%%%%

The part of the disk with radii $R>0.500~{\rm AU}$ has no measurable
influence on the spectrum at wavelength $<5~\mu{\rm m}$. The strong CO
fundamental transition lines ($v=1-0$) in the spectrum are well
reproduced by the model. The weaker $v=2-1$ CO emission lines are also
visible in the spectrum and fitted by our model.  The
\element[][13]{CO} isotope is weakly present in the model, which
assumes the solar system carbon isotope ratio of 89:1, but we cannot
find clear evidence for its existence in the observed spectrum.
\citet{2008arXiv0812.0269W} have modelled isotopic fractionation for a
protoplanetary disk quite similar to our GQ~Lup model, but found only
mild effects on carbon monoxide, with very modest increases in the
\element[][12]{CO}/\element[][13]{CO} ratio only in the outermost
layers of the disk.  Considering the noise in the spectrum and the
blending of \element[][12]{CO} and \element[][13]{CO} lines we
therefore cannot find evidence for a different isotope ratio in our
observations. The maximum temperature in the surface layers of the
%%%%%%
%%%%%%
disk atmospheres is $T=950~{\rm K}$ which is much smaller than the
estimated excitation temperature of $T_{\rm ex}=1814~{\rm K}$ found by
\citet{2003ApJ...589..931N} for DF~Tau which has a similar spectrum
but stronger $v=2-1$ emission lines.

In Fig.~\ref{fig:dust_spec} we show the influence of the dust grain
size distribution on the spectrum. All of the spectra are based on the
same structure model but the emergent flux is calculated for four
different dust grain size setups, i.~e.~no dust, ISM grain size
distribution with $a_0=6.25\cdot 10^{-3}~{\rm \mu m}$, $a_0=6.25\cdot
10^{-2}~{\rm \mu m}$, and $a_0=6.25\cdot 10^{-4}~{\rm \mu m}$. As we
can see, neglecting dust leads to a lack of opacity and too strong
lines. The ISM grain size distrbution fits the data equally well as
the smaller grains. The spectrum with ten times larger grains than ISM
still reproduces the observation reasonably well, but lines are
slightly stronger than for ISM sizes.

An interesting feature in the observed spectrum is the hydrogen
Pf$\beta$ (7$-$5) emission line. This line has been attributed to an
absorption line in the telluric standard star used for the calibration
of CTTS by \citet{2003ApJ...589..931N}. Since we used a telluric model
to remove telluric absorption lines, we can relate the Pf$\beta$
feature unambiguously to an emission line intrinsic to GQ~Lup. We
measure a blue shift of $7~{\rm km/s}$ for the line relative to the
star and the width corresponds to a velocity of $u\approx 172~{\rm
km/s}$. This value is much larger than the measured $v\sin i=6.8~{\rm
km/s}$ by BR07. Therefore, the line cannot be attributed to the star
directly. If we assume Keplerian rotation, the formal origin of the
line is at $R=0.025~{\rm AU}=2~R_\star$ and could originate from the
accretion flow onto the star or a stellar or disk wind.

\section{Summary \& outlook}
We present a new 1D structure and radiative transfer program for
circumstellar disks that is an extension of the general purpose
stellar atmosphere code \texttt{PHOENIX}. The disk is separated into
concentric rings and the structure is calculated from the midplane of
the disk to the top of the disk atmosphere for each ring. A
combination of all rings then yields the total disk spectrum. Our
program assumes a geometrically thin accretion disk geometry and a
critical Reynolds number to describe the energy release in the disk
due to turbulent viscosity. We have modified the hydrostatic equation,
the inner and outer boundary condition of the RT equation, as well as
the energy balance equation to account for the difference between star
and disk atmosphere. Irradiation of the central star onto the top of
the disk atmosphere is taken into account.

Our disk model spectra can reproduce high-resolution IR emission line
spectra, in this case of the CTTS GQ~Lup. For this particular object,
we calculated a set of disk models for different stellar input
parameters given by the recent publications of
\citet{2007A&A...468.1039B} and \citet{2008A&A...489..349S} and varied
the mass accretion rate $\dot{M}$, the Reynolds number $Re$, and the
inner and outer disk radii. We investigated the contribution of each
disk ring on the total disk spectrum and showed where line and
continuum radiation originates in the atmosphere. Furthermore we could
show that the inclination of the warm inner disk of GQ~Lup is $\sim
22^\circ$ which is much smaller than the value obtained by
\citet{2008A&A...489..349S} and just within the uncertainties given
for the inclination by \citet{2007A&A...468.1039B}.

Even though we could provide a reasonable model fit to the observed
spectra of GQ~Lup with our 1D structure and RT code, there is room for
improvement. One obvious drawback of our code is the assumption of
equilibrium chemistry and local thermodynamic equilibrium. This is not
very likely to be valid for the cool inner and optically thin and
irradiated outer disk atmospheric layers. Therefore, the consideration
of departure from chemical equilibrium due to turbulent mixing or
photodissociation, and radiative excitation of molecules needs to be
considered in future projects. Furthermore, convective energy
transport will be considered in the future in order to investigate the
influence of this effect on the disk structure and spectra.

We are extending our effort to model the structure and spectra of
circumstellar disks to full 3D radiative transfer calculations. This
way we can relax the assumption that there is no flux exchange between
matter at different radii $R$ and we will ``naturally'' consider the
direct irradiation of the central star onto the disk and heat the very
inner disk rim. The influence of the Kepler velocity field in the disk
on the line profile will be investigated by means of two level atom
line transfer \citep[see][]{2007A&A...468..255B}.

\section*{Acknowledgements}
S.D.H. is supported by a scholarship of the DFG Graduierten\-kolleg
1351 ``Extrasolar Planets and their Host Stars''. A.S. acknowledges
financial support from the Deutsche Forschungsgemeinschaft under DFG
RE 1664/4-1.

Based on observations made with ESO Telescopes at Paranal Observatory
under programme ID 179.C-0.151(A).

\bibliographystyle{aa}
\bibliography{diskrt}

\begin{thebibliography}{33}
\expandafter\ifx\csname natexlab\endcsname\relax\def\natexlab#1{#1}\fi

\bibitem[{{Allard} {et~al.}(2001){Allard}, {Hauschildt}, {Alexander},
  {Tamanai}, \& {Schweitzer}}]{LimDust}
{Allard}, F., {Hauschildt}, P., {Alexander}, D., {Tamanai}, A., \&
  {Schweitzer}, A. 2001, \apj, 556, 357

\bibitem[{{Balbus} \& {Hawley}(1991)}]{1991ApJ...376..214B}
{Balbus}, S.~A. \& {Hawley}, J.~F. 1991, \apj, 376, 214

\bibitem[{{Baron} \& {Hauschildt}(2007)}]{2007A&A...468..255B}
{Baron}, E. \& {Hauschildt}, P.~H. 2007, \aap, 468, 255

\bibitem[{{Bell} \& {Lin}(1994)}]{1994ApJ...427..987B}
{Bell}, K.~R. \& {Lin}, D.~N.~C. 1994, \apj, 427, 987

\bibitem[{{Broeg} {et~al.}(2007){Broeg}, {Schmidt}, {Guenther}, {Gaedke},
  {Bedalov}, {Neuh{\"a}user}, \& {Walter}}]{2007A&A...468.1039B}
{Broeg}, C., {Schmidt}, T.~O.~B., {Guenther}, E., {et~al.} 2007, \aap, 468,
  1039

\bibitem[{{Cannizzo} {et~al.}(1982){Cannizzo}, {Ghosh}, \&
  {Wheeler}}]{1982ApJ...260L..83C}
{Cannizzo}, J.~K., {Ghosh}, P., \& {Wheeler}, J.~C. 1982, \apjl, 260, L83

\bibitem[{{Carr}(2007)}]{2007IAUS..243..135C}
{Carr}, J.~S. 2007, in IAU Symposium, Vol. 243, IAU Symposium, ed. J.~{Bouvier}
  \& I.~{Appenzeller}, 135--146

\bibitem[{{D'Alessio} {et~al.}(1998){D'Alessio}, {Canto}, {Calvet}, \&
  {Lizano}}]{1998ApJ...500..411D}
{D'Alessio}, P., {Canto}, J., {Calvet}, N., \& {Lizano}, S. 1998, \apj, 500,
  411

\bibitem[{{Dullemond} {et~al.}(2007){Dullemond}, {Hollenbach}, {Kamp}, \&
  {D'Alessio}}]{2007prpl.conf..555D}
{Dullemond}, C.~P., {Hollenbach}, D., {Kamp}, I., \& {D'Alessio}, P. 2007, in
  Protostars and Planets V, ed. B.~{Reipurth}, D.~{Jewitt}, \& K.~{Keil},
  555--572

\bibitem[{{Ferguson} {et~al.}(2005){Ferguson}, {Alexander}, {Allard}, {Barman},
  {Bodnarik}, {Hauschildt}, {Heffner-Wong}, \& {Tamanai}}]{2005ApJ...623..585F}
{Ferguson}, J.~W., {Alexander}, D.~R., {Allard}, F., {et~al.} 2005, \apj, 623,
  585

\bibitem[{{Gammie}(1996)}]{1996ApJ...457..355G}
{Gammie}, C.~F. 1996, \apj, 457, 355

\bibitem[{Grevesse \& Noels(1993)}]{gn93}
Grevesse, N. \& Noels. 1993, in Abundances, ed. C.~Jaschek \& M.~Jaschek
  (Dordrecht: Kluwer), 111

\bibitem[{Hauschildt \& Baron(1999)}]{jcam}
Hauschildt, P.~H. \& Baron, E. 1999, Journal of Computational and Applied
  Mathematics, 109, 41

\bibitem[{Hauschildt {et~al.}(1994)Hauschildt, St{\"o}rzer, \& Baron}]{aliperf}
Hauschildt, P.~H., St{\"o}rzer, H., \& Baron, E. 1994, JQSRT, 51, 875

\bibitem[{{Helling} {et~al.}(2008){Helling}, {Ackerman}, {Allard}, {Dehn},
  {Hauschildt}, {Homeier}, {Lodders}, {Marley}, {Rietmeijer}, {Tsuji}, \&
  {Woitke}}]{2008MNRAS.391.1854H}
{Helling}, C., {Ackerman}, A., {Allard}, F., {et~al.} 2008, \mnras, 391, 1854

\bibitem[{{Horne} \& {Marsh}(1986)}]{1986MNRAS.218..761H}
{Horne}, K. \& {Marsh}, T.~R. 1986, \mnras, 218, 761

\bibitem[{{Hubeny}(1990)}]{1990ApJ...351..632H}
{Hubeny}, I. 1990, ApJ, 351, 632

\bibitem[{{Janson} {et~al.}(2006){Janson}, {Brandner}, {Henning}, \&
  {Zinnecker}}]{2006A&A...453..609J}
{Janson}, M., {Brandner}, W., {Henning}, T., \& {Zinnecker}, H. 2006, \aap,
  453, 609

\bibitem[{{Kaeufl} {et~al.}(2004){Kaeufl}, {Ballester}, {Biereichel},
  {Delabre}, {Donaldson}, {Dorn}, {Fedrigo}, {Finger}, {Fischer}, {Franza},
  {Gojak}, {Huster}, {Jung}, {Lizon}, {Mehrgan}, {Meyer}, {Moorwood}, {Pirard},
  {Paufique}, {Pozna}, {Siebenmorgen}, {Silber}, {Stegmeier}, \&
  {Wegerer}}]{2004SPIE.5492.1218K}
{Kaeufl}, H.-U., {Ballester}, P., {Biereichel}, P., {et~al.} 2004, in Society
  of Photo-Optical Instrumentation Engineers (SPIE) Conference Series, Vol.
  5492, Society of Photo-Optical Instrumentation Engineers (SPIE) Conference
  Series, ed. A.~F.~M. {Moorwood} \& M.~{Iye}, 1218--1227

\bibitem[{{Kenyon} \& {Hartmann}(1995)}]{1995ApJS..101..117K}
{Kenyon}, S.~J. \& {Hartmann}, L. 1995, \apjs, 101, 117

\bibitem[{{Klahr}(2007)}]{2007IAUS..239..405K}
{Klahr}, H. 2007, in IAU Symposium, Vol. 239, IAU Symposium, ed. F.~{Kupka},
  I.~{Roxburgh}, \& K.~{Chan}, 405--416

\bibitem[{{Kriz} \& {Hubeny}(1986)}]{1986BAICz..37..129K}
{Kriz}, S. \& {Hubeny}, I. 1986, Bulletin of the Astronomical Institutes of
  Czechoslovakia, 37, 129

\bibitem[{{Lucy}(1964)}]{1964SAOSR.167...93L}
{Lucy}, L.~B. 1964, SAO Special Report, 167, 93

\bibitem[{{Lynden-Bell} \& {Pringle}(1974)}]{1974MNRAS.168..603L}
{Lynden-Bell}, D. \& {Pringle}, J.~E. 1974, \mnras, 168, 603

\bibitem[{{Meyer} \& {Meyer-Hofmeister}(1982)}]{1982A&A...106...34M}
{Meyer}, F. \& {Meyer-Hofmeister}, E. 1982, \aap, 106, 34

\bibitem[{{Najita} {et~al.}(2003){Najita}, {Carr}, \&
  {Mathieu}}]{2003ApJ...589..931N}
{Najita}, J., {Carr}, J.~S., \& {Mathieu}, R.~D. 2003, \apj, 589, 931

\bibitem[{{Neuh{\"a}user} {et~al.}(2005){Neuh{\"a}user}, {Guenther},
  {Wuchterl}, {Mugrauer}, {Bedalov}, \& {Hauschildt}}]{2005A&A...435L..13N}
{Neuh{\"a}user}, R., {Guenther}, E.~W., {Wuchterl}, G., {et~al.} 2005, \aap,
  435, L13

\bibitem[{{Pontoppidan} {et~al.}(2008){Pontoppidan}, {Blake}, {van Dishoeck},
  {Smette}, {Ireland}, \& {Brown}}]{2008ApJ...684.1323P}
{Pontoppidan}, K.~M., {Blake}, G.~A., {van Dishoeck}, E.~F., {et~al.} 2008,
  \apj, 684, 1323

\bibitem[{{Seperuelo Duarte} {et~al.}(2008){Seperuelo Duarte}, {Alencar},
  {Batalha}, \& {Lopes}}]{2008A&A...489..349S}
{Seperuelo Duarte}, E., {Alencar}, S.~H.~P., {Batalha}, C., \& {Lopes}, D.
  2008, \aap, 489, 349

\bibitem[{{Shakura} \& {Syunyaev}(1973)}]{1973A&A....24..337S}
{Shakura}, N.~I. \& {Syunyaev}, R.~A. 1973, A\&A, 24, 337

\bibitem[{{Shaviv} \& {Wehrse}(1986)}]{1986A&A...159L...5S}
{Shaviv}, G. \& {Wehrse}, R. 1986, \aap, 159, L5

\bibitem[{{Wolf}(2001)}]{2001PhDT........11W}
{Wolf}, S. 2001, PhD thesis, AA(University of Jena)

\bibitem[{{Woods} \& {Willacy}(2009)}]{2008arXiv0812.0269W}
{Woods}, P.~M. \& {Willacy}, K. 2009, \apj, accepted

\end{thebibliography}

\end{document}